\documentclass[12pt]{article}
\usepackage{epsfig,latexsym,amssymb}
\begin{document}

\begin{center}
{\large \bf  ISCO and Principal Null Congruences in Extremal Kerr Spacetime }
\end{center}

\vskip 5mm

\begin{center}
{\Large{Partha Pratim Pradhan\footnote{E-mail: pppradhan77@gmail.com,~ pppradhan@vsm.org.in}}}
\end{center}

\vskip  0.5 cm

{\centerline{\it Department of Physics}}
{\centerline{\it Vivekananda Satabarshiki Mahavidyalaya}}
{\centerline{\it Manikpara, Paschim Medinipur}}
{\centerline{\it West Bengal~721513, India}}

\vskip 1cm


\begin{abstract}
 The effective potential in universal like coordinates$(U,~V,~\theta,~\phi)$, which are smooth across the event horizon is derived  and investigated the ISCO(Innermost Stable Circular Orbits) explicitly in these coordinates for extremal Kerr spacetime. Extremization of the effective potential for timelike circular orbit shows that the existence of a stable circular geodesics in the extremal spacetime for direct orbit, precisely {\it on} the event horizon in terms of the radial coordinate which coincides with the \emph{principal null geodesic congruences} of the event horizon. These null geodesic congruences mold themselves to the spacetime curvature in such a way that Weyl conformal tensor and its dual are vanished, that is why they are in fact \emph{doubly degenerate principal null congruences}.
\end{abstract}




\section{Introduction:}

Generic non extremal black hole spacetime is a different class than the extremal spacetime, it has been proved by several authors for Reissner Nordstr\"om(RN) blackhole\cite{pp1,pp3} and Kerr blackhole\cite{pp2}. There are certain number of features which has been present in the non extremal case but in the extremal case they are completely absent. For example the `bifurcate 2-sphere' which exists in near extremal spacetime to compare the `macroscopic' entropy of a class of extremal black holes solutions of the supergravity limit of string theories, to the `microscopic' entropy obtained from counting of string states\cite{sen} but the extremal spacetimes does not have any such `bifurcate 2-sphere'.

It is well known that extremal spacetime do not have any trapped surface\cite{pp1,pp2} inside the event horizon (itself usually a marginal outer trapped surface ), whereas the near extremal black hole spacetime have filled up with the trapped surface.
Indeed, it is also true that\cite{bek} the proper spatial distance  on a constant time slices from  event horizon to Cauchy horizon in the extremal Kerr geometry is actually infinity but in the near extremal situation the proper distance is finite.
It has been found that\cite{pp2} in the extremal geometry, there exists a class of stable circular orbit so called ISCO on the event horizon of the extremal spacetime which coincides with the null geodesic generator of the event horizon\cite{pp1,pp2} but the near extremal black hole spacetime does not admit this class of geodesics. Lastly, the extremal black hole spacetime have zero Hawking temperature due to the zero surface gravity which measures the equilibrium temperature for the thermal distribution of the radiation, whereas the non extremal blackhole possesses Hawking temperature.

In this \emph{letter}, we show an interesting feature of the {\it Kerr} spacetime, is
that the effective potential for timelike circular geodesics and null circular geodesics \emph{coalesce into zero energy geodesics} and therefore direct ISCO coalesce with the \emph{principal null geodesic congruences}. This happens only in extremal spacetime, near extremal spacetime does not admit such type of features. In particular, the class of geodesics close to the event horizon in the extremal case does not overlap with the class in the case infinitesimally close to the extremal.

The plan of the paper is as follows: In section 2 we explicitly derive the effective potential in universal like coordinates and discuss the subtleties of the coordinates on the horizon. In section 3 we calculate the principal null geodesic congruences . Finally, we conclude our discussions in section 5.

Consider the Kerr metric in Boyer-Lindquist coordinates,

\begin{eqnarray}
ds^2 &=& -\frac{\Delta}{\rho^2} \, \left[dt-a\sin^2\theta d\phi \right]^2+\frac{\sin^2\theta}{\rho^2} \,
\left[(r^2+a^2) \,d\phi-a dt\right]^2
+\rho^2 \, \left[\frac{dr^2}{\Delta}+d\theta^2\right] ~.\label{nkm}
\end{eqnarray}
where $a \equiv \frac{J}{M}$, $\rho^2 \equiv r^2+a^2\cos^2\theta$
$\Delta \equiv r^2-2Mr+a^2\equiv(r-r_{+})(r-r_{-})$.  The horizon occurs at  $r_{\pm}=M\pm\sqrt{M^2-a^2} $. The outer horizon $r_{+}$ is called event horizon and the inner horizon $r_{-}$ is called Cauchy horizon.

\section{Zero Energy Geodesics:}

The extremal Kerr metric is adapted from the work \cite{pp2}( See for details: arXiv:1108.2333v1). Therefore on the equatorial plane $\theta=\pi/2$ the metric can be written as
\begin{eqnarray}
ds^2 &=& {\cal A}\left(\sec^4U\, dU^2+\csc^4V\,dV^2 \right)+{\cal B}\sec^2U \, \csc^2VdUdV +{\cal C}\,(d\phi^{\star})^2 \nonumber \\[4mm] &&
+{\cal D}\left(\sec^2U\, dU+\csc^2V\, dV\right)d\phi^{\star} ~.\label{metex}
\end{eqnarray}
where
\begin{eqnarray}
{\cal A}&=&\frac{1}{8}(1-\frac{M}{r})^2 \left(\frac{r^2}{r^2+M^2}+\frac{1}{2}\right) \left(\frac{r^2-M^2}{r^2+M^2}\right)+\frac{(M^2-r^2)^2}{16M^2r^2} ~.\label{a}\\
{\cal B}&=&\frac{(M^2-r^2)^2}{8M^2r^2}-\frac{1}{2}(1-\frac{M}{r})^2 \, \left[\frac{r^4}
{(r^2+M^2)^2}+\frac{1}{4}\right] ~.\label{b}\\
{\cal C}&=&\frac{(r^2+M^2)^2}{r^2}-M^2 \, (1-\frac{M}{r})^2 \,,
{\cal D}=\frac{1}{2}M \, (1-\frac{M}{r})^2-\frac{(M^4-r^4)}{2Mr^2} ~.\label{d}
\end{eqnarray}
The spacetime (\ref{metex}) has a timelike isometry generated by the timelike
Killing vector field ${\bf \xi}$ whose projection along the 4-velocity ${\bf u}$ of geodesics: $\xi \cdot {\bf u} = -{\cal E}$, is conserved along such geodesics. Now,
${\bf \xi}$ has non-vanishing components $\xi^U, \xi^V$ which can be easily derived
from the fact that in the Schwarzschild coordinate basis ${\bf \xi} =
{\bf \partial}_t$. One obtains $\xi^U= \cos^2 U~,~\xi^V = \sin^2 V$.
Thus, in this coordinate chart, ${\cal E}$ can be expressed as
${\cal E}=-({\cal A}+{\cal B}/2) \, \left[\sec^2U \, u^U+\csc^2V \, u^V \right]+2{\cal D} \, u^{\phi}$.
There is also the `angular momentum' $L \equiv \zeta \cdot {\bf u}$ (where $\zeta
\equiv \partial_{\phi}$) which is similarly conserved. It can be expressed as
$L = {\cal D} \, \left[\sec^2U \, u^U+\csc^2V \, u^V\right]+{\cal C} \, u^{\phi}$.
From normalization condition
\begin{eqnarray}
u^2 &=& {\cal A} \, \left[\sec^4U \, (u^U)^2+\csc^4V \, (u^V)^2\right]+{\cal B} \, \sec^2U \, \csc^2V \, u^U \, u^V   \nonumber \\[4mm] &&
+{\cal D} \, \left[\sec^2U \, u^U+\csc^2V \, u^V \right] \, u^{\phi}+{\cal C} \, (u^{\phi})^2~. \label{u2u2}
\end{eqnarray}

For  {\it circular} geodesics; the radial component of the 4-velocity vanishes:
$u^r=0$. In terms of components in our chosen coordinate chart, this
translates into $r_U u^U + r_V u^V =0$ where $r_U \equiv \partial r / \partial
U$ etc. So that one can obtain for circular geodesics
\begin{eqnarray}
{u^U \over u^V} &=& \cos^2 U  \,  \csc^2 V ~. \label{uux}\\
{\cal E} &=& -\left [2({\cal A}+{\cal B}/2) \, \sec^2U \, u^U+2{\cal D}\, u^{\phi}\right]~.\label{enm}\\
L &=& 2{\cal D} \, \sec^2U \, u^U+{\cal C} \, u^{\phi} ~.\label{metld}
\end{eqnarray}

From normalization of the geodesics equation we get
\begin{eqnarray}
{\bf u}^2 &=& -(u^U \, \sec^2U) \,  {\cal E} +{\cal C} \, (u^{\phi})^2 .~ \label{frvt}
\end{eqnarray}

For timelike circular geodesics ${\bf u}^2=-1$, Using (\ref{enm},~\ref{metld}) one can obtain the energy equation for timelike circular geodesics is
\begin{eqnarray}
\alpha \, {\cal E}^2+\beta \, \cal{E}+\gamma &=& 0  ~.\label{e2}
\end{eqnarray}
where
\begin{eqnarray}
\alpha = G\,{\cal C}\,{\cal D}^2-{\cal C}\,{\cal H}^2, ~~
\beta =2{\cal C}\,{\cal D}GL\,({\cal A}+{\cal B}/2)-2{\cal D}L{\cal H}^2 \nonumber\\
\gamma = {\cal C}\,GL^2 \, ({\cal A}+{\cal B}/2)^2-G \, {\cal H}^2 ,~~
G = 4{\cal D}^2-2{\cal C} \, ({\cal A}+{\cal B}/2),~~
{\cal H} = {\cal C} \, ({\cal A}+{\cal B}/2)-4{\cal D}^2  ~.\label{gh}
\end{eqnarray}

Therefore the effective potential for timelike circular geodesics may be
written as
\begin{eqnarray}
{\cal E} = ({\cal V}_{eff})_{Horizon} = \frac{-\beta+\sqrt{{{\beta}^2-4\alpha\gamma}}}{2\alpha}
~.\label{vef}
\end{eqnarray}

Similarly the effective potential for null circular geodesics can be written as
\begin{eqnarray}
{\cal E} = ({\cal U}_{eff})_{Horizon} = \frac{-\beta+\sqrt{{{\beta}^2-4\alpha\gamma_{0}}}}{2\alpha} ~.\label{uef}
\end{eqnarray}
where $\gamma_{0} = {\cal C}\,GL^2\,({\cal A}+{\cal B}/2)^2$

Observe that the future horizon of the spacetime is given by $U=\pi/2$ with $V$ arbitrary: in other words $r(\pi/2, V)=M$, which is the direct ISCO for extremal Kerr blackhole also.   One can compute the derivatives,  it turns out that $r_U (\pi/2,V)= 2M^2$, while the other derivative of $r$ is regular on the horizon. Now on the future horizon or at $r=M$(direct ISCO)
\begin{eqnarray}
{\cal A}\rightarrow0,~~{\cal B}\rightarrow 0,~~{\cal C}\rightarrow 4M^2,~~ {\cal D}\rightarrow 0,~~
{\alpha}\rightarrow 0,~~{\beta} \rightarrow 0,~~{\gamma} \rightarrow 0,~~G \rightarrow 0,~~ {\cal H} \rightarrow 0 \nonumber\\
{\cal E} \rightarrow 0 ,~~
{u^U \over u^V} = \cos^2 U \csc^2 V \rightarrow 0  ~.\label{limt}
\end{eqnarray}
which implies  $u^U\rightarrow 0$ and $u^V\rightarrow \infty$ on the horizon. It follows that $L$ is a finite quantity which indeed does  not vanish on the horizon.
It can also easily seen that the effective potential for timelike circular geodesics and
null circular geodesics \emph{coalesce into zero energy geodesics.} i.e.
\begin{eqnarray}
{\cal E} = ({\cal V}_{eff})_{Horizon}=({\cal U}_{eff})_{Horizon}\rightarrow 0  ~.\label{zeroE}
\end{eqnarray}
Since both massive particles and photons are coincident to the zero energy geodesics, therefore they must mold themselves to the spacetime curvature. Thus  Weyl conformal tensor and its dual are vanished on that curvature, so they are called doubly degenerate principal null congruences.

\section{Principal Null Congruences:}

In Boyer-Lindquist coordinates the principal null congruences are

\begin{eqnarray}
k^{t} &\equiv & \frac{dt}{d\lambda}=\frac{(r^{2}+a^2)E}{\Delta} \nonumber\\
k^{r} &\equiv & \frac{dr}{d\lambda}=\pm E \nonumber\\
k^{\theta} &\equiv& \frac{d\theta}{d\lambda}=0 \nonumber\\
k^{\phi}&\equiv& \frac{d\phi}{d\lambda}=\frac{aE}{\Delta} ~.\label{pnc}
\end{eqnarray}
These photon geodesics mold themselves to the spacetime curvature in such a way that, if
$C_{\alpha\beta\gamma\delta}$ is the Weyl conformal tensor and ${\ast}C_{\alpha\beta\gamma\delta}=\epsilon_{\alpha\beta\mu\nu}C^{\mu\nu}_{\gamma\delta}$
is its dual, then

\begin{eqnarray}
C_{\alpha\beta\gamma}[{\delta} k_{\epsilon}]k^{\beta}k^{\nu}=0\, ,
{\ast}C_{\alpha\beta\gamma }[{\delta} k_{\epsilon}]k^{\beta}k^{\nu}=0  ~.\label{weyl}
\end{eqnarray}
This implies, these photon geodesics are doubly degenerate principal null congruences.

The fact that the proper spatial distance\cite{pp2} from ISCO to horizon tends to non zero  and diverges at the extremal limit. But the proper time interval for in-falling radial geodesics for zero angular momentum on the horizon at the extremal limit will
be $\frac{dr}{d\tau}= \pm \sqrt{\frac{2M}{r}(1+\frac{a^2}{r^2})}$. Now at the horizon at the extremal limit this can be $\frac{dr}{d\tau}= \pm 2$. $+$ sign for outgoing geodesics and $-$ sign for ingoing geodesics. Therefore in-falling  proper time from photon orbit or ISCO or marginal bound orbit  to the horizon becomes zero. This implies the fact that these orbits both coincides to the principal null congruences of the future event horizon.

\section{Discussion:}

Therefore the issue of direct ISCO on the horizon  must \emph{coincides} with the principal null geodesic congruences of the event horizon. Thus the effective potential for timelike circular geodesics and null circular geodesics \emph{coalesce into zero energy geodesics} . This zero energy geodesics are hovering on the horizon but does not fall in or out.  They eventually mold themselves to the spacetime curvature. This happens only in extremal blackhole spacetime.  The possible existence of this zero energy trajectory on the event horizon of an extremal Kerr spacetime is a surprising feature. The evidence of this is that it appears as a global minimum of the effective potential. Notice that, the location of the ISCO on the event horizon is clearly indicated by the effective potential, its actual existence may illusory\cite{bpt}, since generically the black hole event horizon has a spatial foliation as a marginal outer trapped surface. Further, it is the limit of an infinite set of outer trapped surface `inside' it i.e. for $r<M$. This implies that any stable circular orbit of massive particle and null circular orbit of massless particles on and encircling the horizon must end up with the null geodesic generators of the future horizon.

\vglue .5cm

\end{document}